\documentclass[a4paper,12pt]{article}
\usepackage{amsfonts,amsthm,amsmath,amssymb,graphicx,hyperref,youngtab}
\usepackage{enumerate}
\usepackage{amsmath,amsfonts,dsfont}
\usepackage{amssymb}
\usepackage{tikz}
\usetikzlibrary{intersections, angles, quotes}
\usetikzlibrary{positioning}
\usepackage{tikz-cd}
\usetikzlibrary{arrows,decorations.markings}
\usetikzlibrary{quotes, calc}
\usepackage[T1]{fontenc}
\usepackage{authblk}
\usepackage[latin9]{inputenc}
\usepackage{color}
\usepackage{latexsym}
\usepackage{amsbsy}
\usepackage{amstext}
\usepackage{graphicx}
\usepackage{amsmath}
\usepackage{amssymb}
\usepackage{tikz}
\usepackage{mathtools}
\usepackage{amsthm}
\usepackage{amstext}
\usepackage{amssymb}
\usepackage{graphicx}
\usepackage{esint}
\usepackage[english]{babel} 

\usepackage{times}
\usepackage{color}
\usepackage{fancyhdr}
\newcommand{\cA}{\mathcal A}

\newcommand{\cC}{\mathcal C}

\newcommand{\cF}{\mathcal F}

\newcommand{\cH}{\mathcal H}

\newcommand{\cL}{\mathcal L}
\newcommand{\cM}{\mathcal M}
\newcommand{\cN}{\mathcal N}
\newcommand{\cO}{\mathcal O}

\newcommand{\cZ}{\mathcal Z}

\newcommand{\be}{\begin{equation}}
\newcommand{\ee}{\end{equation}}
\newcommand{\bea}{\begin{eqnarray}}
\newcommand{\eea}{\end{eqnarray}}

\date{}
\begin{document}
\title{Toric Quiver Asymptotics and Mahler Measure: $\cN =2$ BPS States}
\author[1,2 \footnote{E-mail: a.zahabi@qmul.ac.uk}]{Ali Zahabi}
\affil[1]{\footnotesize Centre for Research in String Theory, School of Physics and Astronomy, Queen Mary University of London, UK}
\affil[2]{National Institute for Theoretical Physics,
School of Physics and Mandelstam Institute for Theoretical Physics,
University of the Witwatersrand, Johannesburg, SA}
\maketitle
\abstract
BPS sector in $\mathcal{N}=2$, four-dimensional toric quiver gauge theories has previously been studied using crystal melting model and dimer model. We introduce the Mahler measure associated to statistical dimer model to study large $N$ limit of these quiver gauge theories.
In this limit, generating function of BPS states in a general toric quiver theory is studied and entropy, growth rate of BPS states and free energy of the quiver are obtained in terms of the Mahler measure. Moreover, a possible third-order phase transition in toric quivers is discussed.
Explicit computations of profile function, entropy density, BPS growth rate and phase structure of quivers are presented in concrete examples of clover $\mathbb{C}^3$, and resolved conifold $\mathcal{C}$ quivers. Finally, BPS growth rates of Hirzebruch $\mathbb{F}_0$, and $\mathbb{C}^3/ \mathbb{Z}^2\times \mathbb{Z}^2$ orbifold quivers are obtained and a possible interpretation of the results for certain BPS black holes is discussed. 

\maketitle

\section{Introduction}
In the context of AdS/CFT correspondence, quiver gauge theories are engineered by low energy limit of D-branes on singularities of Calabi-Yau manifolds \cite{Do-Mo}. In this way, half-BPS states in $\cN=2$ four-dimensional quiver gauge theories can be seen as D6-D2-D0 branes bound states in type IIA string theory. Recently, the BPS sector, its rich structure and interesting features such as wall-crossing and relations to topological strings are studied extensively, \cite{Ko-So, Ga-Mo, Oo-St}. Beside AdS/CFT, quiver representation theory and combinatorial lattice models constructed from that, such as dimer model and crystal model play an important role in study of non-perturbative aspects of quiver gauge theories \cite{Ya-Oo2}. 

Despite intense studies about the BPS counting and interesting behavior of BPS index, the asymptotic analysis of the index is less considered, while, some interesting features of the theory such as phase transitions appear in the asymptotic limit. The main goal of this paper is to study the half BPS sector of $\cN=2$, $4d$ toric quiver gauge theories
on the universal cover of quiver diagram, and to obtain large $N$ asymptotic of the BPS index.
Similar problems in the asymptotic limit of different gauge theories have been studied using other plausible methods. Recently, asymptotic counting of chiral operators in free quiver gauge theory is studied via multivariate asymptotic analysis of generating functions \cite{Ra-Wi-Za}. In another direction, phase structure of a topological gauge theory, Chern-Simons matter theory is studied by using random matrix theory \cite{Za}.

Quiver lattice models and related mathematical techniques, provide efficient methods for counting BPS states and its asymptotic limit. Crystal melting model captures the BPS index of toric quiver gauge theories. The crystal melting model is also interesting from the asymptotic point of view. In fact, thermodynamic limit of the crystal model is studied via toroidal dimer model \cite{Ke-Ok-Sh}. Therefore, statistical mechanics of the dimer model on a toroidal bipartite graph and related techniques from tropical geometry and Mahler measure theory provide a powerful framework to study the asymptotics of BPS sector of toric quiver gauge theories. Furthermore, 
we use generalized Mahler measures (on arbitrary tori) of certain Newton polynomials related to particular toric Calabi-Yau threefolds, $\mathbb{C}^3$ and resolved conifold, to explicitly compute some analytic results in the quiver theory.

Using analytic techniques from \cite{Ke-Ok-Sh}, thermodynamic limit of the generating function of BPS states is studied via the dimer model and the free energy of the quiver theory are obtained in \cite{Ya-Oo}.
Following this approach and adopting methods from statistical toroidal dimer model and associated Mahler measure theory in the quiver gauge theory, we obtain parameters and observables of the quiver theory, such as profile of rank of gauge groups of quiver, free energy, entropy, BPS growth rate and phase structure of the theory, in the asymptotic limit.

In explicit computations in concrete examples of quivers, we use dilogarithm function and its associated special functions in Mahler measure theory. Using Lobachevsky and Bloch-Wigner functions, we develop statistical mechanics of the clover and resolved conifold quivers in the thermodynamic limit. We explicitly compute the observables of the these quiver gauge theories and obtained new analytic expressions for them.

One of our first observations of this work is that the BPS growth rate of any toric quiver can be obtained, up to a scaling, from the surface tension of the profile function and in terms of the Mahler measure of the Calabi-Yau threefold. The actual computations of the BPS growth rate is performed explicitly in most important and fundamental examples, such as $\mathbb{C}^3$, resolved conifold $\mathcal{C}$, local $\mathbb{P}^1 \times \mathbb{P}^1$, and orbifold quivers. 

The free energy of quiver is shown to be equal to the partition function of the genus zero topological strings \cite{Ya-Oo}. Following this result, we compute the finite part of the topological string partition function, in certain approximations, in terms of Mahler measure and toric data.

To study phase structure of the toric quiver gauge theory, we use the tropical geometric methods in toroidal dimer model \cite{Ke-Ok-Sh}. Similar to dimer model, the phase structure of the quiver is obtained from different fluctuating behavior of the profile function in different regions of the Amoeba. Using the entropy density of quiver, we find evidences for a possible third order phase transition at the boundaries of Amoeba.

The crystal melting model for $\mathbb{C}^3/ \mathbb{Z}^p \times \mathbb{Z}^p$ orbifold quiver in large $p$ limit, explains certain BPS black holes \cite{He-Va}. We propose physical applications and possible interpretations of our results for BPS black holes in this example of large orbifold quivers.

This paper is organized as follow. In section 2, we review the crystal melting model for BPS states of the toric quiver theories. In section 3, we first briefly review some backgrounds in statistical dimer model and its connection to Mahler measure theory and special functions, then thermodynamics of the quiver theories are explained via the Mahler measure. In section 4, explicit computations in some concrete examples are presented. In section 5, we conclude our paper and discuss some directions for future research. 

\section{BPS Quivers and Crystal Model}
\label{sec:quiv crys}
In this part, we briefly review the BPS bound states of D6-D2-D0 branes on a toric Calabi-Yau threefold singularity and its relation to crystal melting model \cite{Ya-Oo2}. We consider a single non-compact D6-brane wrapping the whole Calabi-Yau threefold and arbitrary number of point-like D0 branes, and D2-branes wrapping compact two-cycles in the Calabi-Yau.  

The D-brane bound states are $\frac{1}{2}$-BPS states in $\cN=2$, $4d$ gauge theories. They carry a charge $\gamma=(n,l)$ where $n$ and $l$ are the number of D0 and D2 branes, respectively. The number of BPS states with a given charge $\gamma$ is denoted by $\Omega(\gamma)$ and is called BPS index. Defining the fugacity factors $q$ and $Q_I$ for D0 and D2 charges, the generating function of BPS index is given by
\bea
\label{BPS gf}
\cZ_{BPS}=\sum_{n, l} \Omega(n, l) Q^l q^n,
\eea
where $Q=\{Q_I\}$ and $Q^l=\prod_{I} Q_I^{l_I}$ with $I$ runs over two-cycles in the Calabi-Yau. Writing $q=e^{-g_s}$ and $Q_I=e^{t_I}$, we can identify $g_s$ with the coupling constant of topological string and $t_I$ as the K\"{a}hler moduli of the Calabi-Yau threefold.

In order to compute the index, we consider the low energy effective theory on the D-branes, which is a $d=1$, $\cN=4$ toric quiver quantum mechanics. In this way, the BPS index reduces to the Witten index. Moreover, field content and the superpotential term of quantum mechanics is encoded in the quiver diagram. Quiver diagram $Q=(Q_0, Q_1)$ consists of set of nodes $Q_0$ with a gauge group $U(N_i)$ associated to each node $i$, and set of arrows $Q_1$ connecting the nodes with  a bifundamental field associated to each arrow. The set of faces of quiver is denoted by $Q_2$.  The dual diagram of quiver diagram is denoted by $Q'$ and is called brane tiling. Quiver diagram is related to brane tiling by the following dictionary: $Q_0'=Q_2$, $Q_1'=Q_1$, and $Q_2'=Q_0$. Superpotential $W$ consists of terms which are represented as a loop around a face in the quiver diagram. For further details of how D0, D2 and D6 branes appear in this picture, see \cite{Ya-Oo2}. Assigning a black or white color to each vertex of $Q_0'$ depending on the orientation of the corresponding face, the brane tiling defines a toroidal bipartite dimer model \cite{Ke-Ok-Sh}. The statistical mechanics of this dimer model plays a crucial role in this study.

To compute the Witten index, we need to identify the moduli space of vacua and this can be done in terms of quiver representation. There is a path algebra $\mathbb{C}Q$ associated to a quiver diagram, with paths between nodes of the quiver as elements and concatenation of paths as product. Imposing F-term relations on path algebra, we obtain a factor algebra as a quotient, $A= \mathbb{C}Q/F$, where $F$ is the ideal of path algebra generated by $\partial W/ \partial X_a$ for $X_a \in Q_1$. This is often called A-module. Furthermore, imposing D-term relations, we get a so-called stable A-moldule which is in one-to-one correspondence with supersymmetric vacua of quiver quantum mechanics.

Consider the universal cover of quiver diagram $\Tilde{Q}$. Any path starting at a reference node $i_0\in \Tilde{Q}_0$ and ending at $j\in \Tilde{Q}_0$ in the A-module can be given in the form $v_{i_0 j}\ \omega^n$, for some $n\in \mathbb{N}$, where $v_{i_0 j}$ is the shortest path between vertices $i_0$ and $j$ and $\omega$ is a superpotential term, i.e. a loop around some face of $\Tilde{Q}$. The crystal is defined by putting an atom on ending point of each path on the vertices of the quiver. The paths of the form $v_{i_0 j}$ correspond to atoms on the face of the crystal, while paths $v_{i_0 j}\ \omega^n$ with $n>0$ are atoms on the node $j$ at the depth $n$ inside the crystal. Putting atoms for all elements of factor algebra, we construct a three-dimensional crystal on the universal cover of periodic quiver. 

Now, the Witten index can be computed in terms of stable A-module, as a sum over the fixed-points of the torus action, more precisely $U(1)^3\times U(1)_R$, on the moduli space. The fixed-points of the moduli space are equivalent to $U(1)^3$ invariant stable A-modules, which is in one-to-one correspondence with the ideals of factor algebra. The ideals of factor algebra are in fact the molten configurations of crystal. Thus, the fixed points are in bijection with the molten crystal configurations, and generating function of the Witten index, up to a sign, can be written as a sum over molten crystals $\pi_i\in \pi$,
\bea
\label{CM gf}
\cZ_{BPS}=\cZ_{CM}= \sum_{\pi}\prod_i q_i^{|\pi_i|},
\eea
where $q_i$ is a fugacity factor associated to $|\pi_i|$, i.e. the number of atoms in $\pi$ which is associated with the $i$-th quiver node. By comparing generating functions \eqref{BPS gf} and \eqref{CM gf}, we see that the Witten index is equal to the number of molten crystal configurations
where $n$ is the total number of atoms removed from the crystal, and the relative numbers of
different types of atoms removed from the crystal are specified by $l_I$'s.
The main goal of this paper is the asymptotic counting of BPS states, $\Omega(n,l)$ in the large $n$ limit and obtaining the BPS growth rate, i.e. logarithm of BPS index in the asymptotic limit.

\section{Asymptotic Analysis in BPS Sector}
In this chapter we explain a method of asymptotic analysis in the BPS sector of toric quivers. The BPS sector of quiver gauge theories is a quantum mechanical system and statistical mechanics of the crystal model explains this sector in the asymptotic regime. To study the asymptotic regime, we adopt the statistical mechanics of the toroidal dimer model and Mahler measure theory in the context of quiver gauge theories. Before that, we review some background materials.
\subsection{Mahler Measure in Toroidal Dimer Model}
\label{sec:dimer}
In this section, we review essential backgrounds and methods from statistical mechanics of biperiodic bipartite dimer model \cite{Ke-Ok-Sh}, and associated Mahler measure.

Dimer model on a bipartite planar graph is the statistical mechanics of set of dimer coverings of the graph, i.e. a pairing of adjacent vertices such that each vertex is covered only by one dimer. 
Consider a toroidal bipartite graph $G=(V,E)$, as a set of vertices $v\in V$ and edges $e\in E$, with oriented closed curves $\gamma_z$ and $\gamma_w$ on $\mathbb{T}^2$, transverse to $G$, representing a basis of $H_1(\mathbb{T}^2)$. Each edges of the graph is oriented from a black vertex to a white vertex. The weight $\nu(e)$ of edge $e$ is $\nu(e)=z^{\gamma_z \cdot e} w^{\gamma_w\cdot e}$, with $\cdot$ denotes the signed intersection number of edge and closed curve. A Kasteleyn matrix $K(z,w)$ is an oriented adjacency matrix of $G$, indexed by vertices, and the matrix elements of $K$ are $\pm\nu(e)$ with the sign given by the orientation of that edge. Orient the edges of $G$ such that each face of graph has an odd number of clockwise oriented edges.
The characteristic polynomial of the dimer model, called the Newton polynomial, is defined by
$$P(z,w)= \det K(z,w).$$
A fundamental result is the number of the dimer coverings of $G$, obtained in \cite{Ke-Ok-Sh},
$$Z(G)=\frac{1}{2}|-P(1,1)+P(-1,1)+P(1,-1)+P(-1,-1)|.$$ For an integer $M>0$, define a finite graph on a torus, $G_M=G/ M\mathbb{Z}^2$. Then, logarithm of partition function of the toroidal dimer model on the universal cover of the graph, which is the growth rate of the toroidal dimer model on $G_M$, is obtained in \cite{Ke-Ok-Sh},
\bea
\label{dim mah}
\log Z(G)=\lim_{M\rightarrow \infty} \frac{1}{M^2} \log Z(G_M) =\frac{1}{(2\pi \mathrm{i})^2} \int_{S^1\times S^1} \log|P( z, w)|\ \frac{dz}{z}\frac{dw}{w},
\eea
where $S^1\times S^1$ is a unit torus defined by $|z|=1, |w|=1$.
Above integral in Eq.~\eqref{dim mah} is called Mahler measure of Newton polynomial, and is denoted by $m(P)$.
A general logarithmic Mahler measure is defined as an average of logarithm of a given non-zero polynomial $P(z_1,z_2, ..., z_n)\in \mathbb{C}[z_1^{\pm}, z_2^{\pm}, ..., z_n^{\pm}]$ over the real $n$-torus $\mathbb{T}^n\in \mathbb{C}^n$,
\bea
\label{Mahler}
m(P)=\frac{1}{(2\pi \mathrm{i})^n} \int_{\mathbb{T}^n} \log|P(z_1,z_2, ..., z_n)|\prod_{i=1}^n\frac{d z_i}{z_i}.
\eea
A generalization of the two-variable Mahler measure on arbitrary tori,
\bea
m_{a,b}(P)=\frac{1}{(2\pi \mathrm{i})^2} \int_{|z|=a}\int_{|w|=b} \log|P( z, w)|\ \frac{dz}{z}\frac{dw}{w},
\eea
is being considered recently \cite{Lal}.  In this study, we consider two explicit examples of this generalization of Mahler measure and apply them to study the asymptotic limit of quiver gauge theories. Explicit computation of the Mahler measure often involves the dilogarithm function and its extensions such as Lobachevsky and Bloch-Wigner functions. In the last part of this section, we briefly review these functions and their properties.
\subsubsection*{Lobachevsky and Bloch-Wigner Functions}
\label{sec:LBW}
Lobachevsky function $L(\theta)$ is defined by the following integral
\bea
L(\theta)= -\int_0^\theta \log|2 \sin (t)|\ dt.
\eea
This function is a well defined, continuous, odd and $\pi$-periodic function for all $\theta$, \cite{Rat}. For any positive integer $n$, $L(\theta)$ satisfies 
\bea
L(n\theta)=n \sum_{j=0}^{n-1} L(\theta+ \frac{j\pi} {n}).
\eea

Bloch-Wigner function, for $z\in \mathbb{C}\setminus \{0,1\}$ is defined as
\bea
D(z)=\Im (Li_2(z))+\log|z|\arg(1-z),
\eea
where $\Im$ denotes the imaginary part and $Li_2(z)$ is the dilogarithm function. For $z\in \mathbb{R}\cup \{\infty\}$,  $D(z)=0$, and
for imaginary $z=e^{i\theta}$, $\theta\in \mathbb{R}$, we have
\bea
D(e^{\mathrm{i}\theta})=\Im[Li_2(e^{\mathrm{i}\theta})]= 2L(\frac{\theta}{2}).
\eea
Bloch-Wigner function is a real-valued analytic function on $\mathbb{C}$ except at 0 and 1 where it is continuous but not differentiable. It satisfies the following equations called 6-fold symmetry \cite{Zag},
\bea
D(z)=D(1-\frac{1}{z})=D(\frac{1}{1-z})=-D(\frac{1}{z})=-D(1-z)=-D(\frac{-z}{1-z}).
\eea
In addition we have
\bea
D(\bar{z})=-D(z), \quad 2D(z)+2D(-z)= D(z^2).
\eea
The Bloch-Wigner function is related to Lobachevsky function in the following way \cite{Lew}.\quad\quad
For $\zeta\in \mathbb{C}\setminus \mathbb{R}$, $\theta_1=\arg(\zeta)$, $\theta_2=\arg(1-\bar{\zeta})$, $\theta_3=\arg(\frac{1}{1-\zeta})$, we have 
\bea
\label{ent clov loba}
D(\zeta)=L(\theta_1)+ L(\theta_2) + L(\theta_3)= L(\theta_1)+ L(\theta_2) - L(\theta_1+\theta_2).
\eea
Having reviewed essential backgrounds, we apply these methods to quiver gauge theories.
\subsection{Toric Quivers in Asymptotic Limit}
In this part, we will apply asymptotics of dimer model and method of Mahler measure theory to study thermodynamics of quiver gauge theory.

The construction of quiver crystal, discussed in Section~\ref{sec:quiv crys}, implies that profile function of ranks of the gauge groups of the quiver is the height function of the crystal model. In the large $N$ limit, with appropriate scaling of the profile function and locations of quiver nodes, the profile becomes a smooth convex function, called limit shape of the crystal model, and it is given by a generalized Mahler measure (Ronkin function) of the Newton Polynomial associated to the quiver \cite{Ke-Ok-Sh},
\bea
\label{Ronkin}
N(x,y)=\frac{1}{(2\pi \mathrm{i})^2} \int_{S^1\times S^1} \log|P(e^x z,e^y w)|\ \frac{dz}{z}\frac{dw}{w}=m\Big(P(e^x z, e^y w)\Big)=m_{e^x,e^y}(P).
\eea
The slope of the profile is defined by 
\bea
\label{slopes}
\frac{\partial N}{\partial x}= n_x, \quad \frac{\partial N}{\partial y}= n_y.
\eea
The results in \cite{Ke-Ok-Sh} imply that surface tension of the profile is the Legendre transform $\cL[\cdot]$ of the profile of the quiver,
\bea
\label{Leg}
-\sigma (n_x, n_y) = \cL[N] = N(x,y) - n_x\ x - n_y\ y,
\eea
where $n_x$ and $n_y$ are given by Eq.~\eqref{slopes}. The Legendre duality also implies
$\frac{\partial\sigma}{\partial n_x}=x,\ \frac{\partial\sigma}{\partial n_y}=y$. 

The Newton Polygon $NP(P)$ of a polynomial $P(z,w)\in \mathbb{C}[z^\pm, w^\pm]$ is a convex hull of the exponents of the monomials with nonzero coefficients. 
Then, it is shown that the slope $(n_x,n_y)$ of the toroidal dimer model lies inside the Newton polygon, $(n_x,n_y)\in NP(P)$ \cite{Ke-Ok-Sh}.
As we will see in the following, the surface tension of the quiver is entropy density of the BPS states and determines the BPS growth rate.

We consider the toric quiver crystals with crystal lattice spacing $g_s$, on the $M\times M$ cover of the torus on the $(x,y)$ plane. One can also think of $g_s$ as inverse temperature. To study quiver crystals in the continuum and thermodynamic limits, the distance between atoms in crystal are taken to zero; $g_s\rightarrow 0$ and we put the quiver on the universal cover; $M\rightarrow \infty$ such that $g_s M$ is a large but fixed. In this way, the number of atoms removed from crystal is taken to infinity while the volume of the crystal $g_s^3 n$ is large but fixed. We call this limits as asymptotic limit. 

In the asymptotic limit, the generating function of the BPS states can be approximated by a dominant configuration or a typical state, coming from the profile of the quiver, in a continuous integral form in which $M$ and $g_s$ dependence are both manifest \cite{Ya-Oo},
\bea
\label{BPS gf 2}
\cZ_{BPS}\sim\exp{\Bigg(M^2\left(\int_{0}^1 -\sigma(n_x, n_y)\ dx\ dy- g_s M \int_{0}^1 N(x,y)\ dx\ dy\right) \Bigg)}.
\eea
With the choice of scaling $M^3 \int_{0}^1 N(x,y)\ dx\ dy = n$, the volume factor in generating function \eqref{BPS gf}, $q^n=e^{-g_s n}$, is given by the second integral in the right hand side of generating function \eqref{BPS gf 2}. One-to-one correspondence between dimer model and crystal model implies that the number of dimer covering is (up to a multiplicative factor) equal to the number of molten configurations in the crystal model, and number of BPS states in toric quivers. In the asymptotic limit, the first integral in Eq.~\eqref{BPS gf 2} gives the asymptotic number of dimer coverings. In the asymptotic regime, the correspondence holds up to a scaling power. The BPS growth rate is obtained by the number of (fluctuating) stepped surfaces lying close to the limit shape. Since the first integral in generating function \eqref{BPS gf 2}, is a volume term, thus its third root gives the correct dimension and with the choice of scaling $n\sim M^3$, BPS growth rate becomes
\bea
\label{gf ent}
\lim_{n\rightarrow\infty}\log \Omega(n,l) \sim n^{\frac{2}{3}}\left(\int_{0}^1 -\sigma(n_x, n_y)\ dx\  dy\right)^\frac{1}{3}.
\eea
Notice that in the asymptotic limit, $Q^l$ term in the generating function \eqref{BPS gf} is finite and thus asymptotically irrelevant.

\subsection{Entropy Density and BPS Growth Rate}
First, by using Eqs.~\eqref{Ronkin} and \eqref{Leg}, we finds the relations between profile function, entropy density and Mahler measure as
\bea
\label{Mahler Ronkin}
-\sigma(n_x^*, n_y^*)=N(0,0)=m(P),
\eea
where $\sigma(n_x^*, n_y^*)$ is the extremum of entropy density obtained by solving
\bea
\frac{\partial\sigma}{\partial n_x}=x=0,\quad \frac{\partial\sigma}{\partial n_y}=y=0.
\eea

For a better approximation, we maximize the entropy density in Eq.~\eqref{gf ent}, and by using Eq.~\eqref{Mahler Ronkin}, we obtain the BPS growth rate
\bea
\log{\Omega(n)}
\sim \ n^{\frac{2}{3}}\left(\int_{0}^1 -\sigma(n^*_x, n^*_y)\ dx\ dy\right)^\frac{1}{3}
= \  n^{\frac{2}{3}}\left(\int_{0}^1 N(0, 0)\ dx\ dy\right)^\frac{1}{3}.\nonumber\\ 
\eea
We thus obtain the numerical coefficient of the BPS growth rate from the asymptotic rank of the quiver at origin, which is the Mahler measure,
\bea
\label{BPS growth2}
\log{\Omega(n)}\sim \ N(0,0)^{1/3} \ n^{2/3}= \ m_P^{1/3} \ n^{2/3}.
\eea

\subsection{Free Energy of Quivers and Topological Strings}
In the infinite volume limit $g_s M\rightarrow\infty$, free energy of quiver which is defined as logarithm of generating function \eqref{BPS gf}, can be obtained, by using the Legendre transformation \eqref{Leg} and putting the integral of the total derivative to zero \cite{Ya-Oo},
\bea
\cF =\log \cZ_{BPS}\sim  \frac{1}{g_s^2} \int_{\mathbb{R}^2} N(x,y)\ dx\ dy .
\eea
There is only $g_s$ dependence remained in above formula and as it suggests, this free energy matches with the genus zero topological string partition function on toric Calabi-Yau threefolds, as it is shown in \cite{Ya-Oo}.
Moreover, at the origin $x=y=0$, it reproduces Eq.~\eqref{dim mah} and thus counts the number of dimer coverings. The above integral is divergent because of infinite contribution from facets of the profile function in the unbounded compliment components of the Amoeba. For $g_s$ small but fixed, finite part of the free energy is obtained by removing the infinite contributions and thus restricting the domain of integration to the Amoeba,
\bea
\label{fin free energy}
\cF_f \sim \frac{1}{g_s^2} \int_{\cA} N(x,y)\ dx\ dy .
\eea
Using the following results about the area of amoeba, a plausible approximation to the integral can be found.
The area of the Amoeba is bounded from above by the area of the corresponding Newton polygon $\Delta$, \cite{Mi-Ru},
\bea
\label{Am ineq}
Area(\cA) \leq \pi^2 Area(\Delta).
\eea
The Newton polynomials of the toroidal dimer models are real curves of special type, called Harnack curves \cite{Ke-Ok-Sh}. Inequality \eqref{Am ineq} saturates for the Amoeba of the Harnack curves and thus for the Amoeba of the dimer model. 
We approximate the finite part of the free energy \eqref{fin free energy}, at origin by using Eq.~\eqref{Mahler Ronkin}, and above theorems,
\bea
\cF_f\sim \frac{1}{ g_s^2} \int_{\cA} N(0, 0)\ dx\ dy
= \frac{m_P}{ g_s^2}  Area(\cA)
= \frac{\pi^2 m_P}{g_s^2} Area(\Delta).
\eea
This is in fact proportional to the number of dimer coverings inside the Amoeba, which is the liquid phase of the dimer model.
In Section~\ref{sec:examples}, we obtain analytic results for profile function and Mahler measure which can be used to explicitly compute the partition function of genus-zero topological string. However, in this paper we do not continue in this line of research.

\subsection{Phase Structure of Quivers and PDEs}
\label{sec:phase}
As quivers are dual to the brane tilings, thus by construction, the phase structure of the toric quiver theory is the same as the phase structure of the toroidal dimer model. This phase structure is determined by the Amoeba \cite{Ke-Ok-Sh}. Amoeba $\cA(P)$ of a Newton polynomial is the image of the zero set
$\{(z,w)\in \mathbb{C}^2| P(z,w)=0\}$,
under the logarithmic map $(z,w)\rightarrow (\log|z|, \log|w|)$, \cite{Mik}.
There are solid, liquid, and gas phases, according to different behavior of fluctuations of the profile function in different regions of the Amoeba \cite{Ke-Ok-Sh}. In unbounded complement components of the Amoeba which correspond to facets of the crystal, the profile function has no fluctuation and this regions are called solid phase. Liquid phase is the interior of the Amoeba, corresponds to melting part of the crystal in which profile function has bounded fluctuations. Gas phase is bounded component of compliment of the Amoeba and inside this region, the profile function has unbounded fluctuations. Gas phases are associated to the points inside the Newton polygon. The phase borders are the boundaries of the Amoeba between different phases.

From the entropy density point of view, phase transition and its order are obtained from discontinuity of entropy and its derivatives, across the boundaries of the Amoeba. In our case, entropy is defined as a double integral of the entropy density, i.e the first integral in generating function \eqref{BPS gf 2}. Although we are not computing contributions of the fluctuations of the profile in the entropy, but we can study whether the entropy density or its derivatives jumps accross the boundaries of the amoeba, since there are different fluctuation contributions to the entropy density in different regions of the amoeba.  Notice that boundaries of the Amoeba correspond to borders of the facets of crystal in which the profile function is constant. On one facet of quiver and its boundary, slope $(n_x,n_y)$ is zero. Thus, 
phase transition on one of the boundaries of the Amoeba can be studied by considering zero slope limit,  $n_x,n_y\rightarrow 0$, of the entropy density and its derivatives for $n_x,n_y\in NP(P)$.  Heuristically, discontinuity in $\lim_{n_x, n_y \rightarrow 0}\frac{\partial^k\sigma(n_x,n_y)}{\partial n_x^k}$ and/or  $\lim_{n_x, n_y \rightarrow 0}\frac{\partial^k\sigma(n_x,n_y)}{\partial n_y^k}$ across the boundaries of Amoeba suggests a possible phase transition of the order $k+2$, since the entropy is a double integral of entropy density.

Using other plausible methods from statistical mechanics of dimer model, it is conjectured that there is a third order phase transition in a dimer model \cite{Co-Pr}. In this article, we obtain some evidence toward a possible third order phase transition in concrete examples of clover and resolved conifold quivers, in Section \ref{sec:examples}.

\subsubsection*{PDEs for Profile and Entropy Density}
As we explained, we expect that some derivatives of entropy density are divergent and generally not continuous near the boundaries of facets of quiver. However, the second order derivatives of profile function and entropy density satisfy the Monge-Amp\`{e}re differential equation inside the Amoeba and near its boundaries.
Define the Hessian matrix of the Ronkin function, $\cH_{ij}(N)\coloneqq \partial_{i j} N$ and the Monge-Amp\`{e}re operator $\cM(N)\coloneqq\det\cH(N)$.
Using results for Ronkin function in \cite{Mi-Ru, Ke-Ok-Sh}, we observe that the profile function satisfy a Monge-Amp\`{e}re equation,
\bea
\label{Mong Ron}
\cM(N)=\partial_{x x} N\ \partial_{y y} N - \partial_{x y} N\ \partial_{y x} N= \pi^2.\quad
\eea
Legendre duality between the profile function and the entropy density implies \cite{Co-Hi},
\bea
\label{Mong Leg}
\cM(N)\cM(\sigma)=1,\quad \partial_{n_j n_i}\sigma=\frac{\cH(N)_{(i j)}}{\cM(N)},\quad \partial_{j i}N=\frac{\cH(\sigma)_{(i j)}}{\cM(\sigma)},
\eea
where $\cH_{(i j)}$ is the co-factor of $i$-th row and $j$-th column of the Hessian matrix. Then, Eqs. \eqref{Mong Ron} and \eqref{Mong Leg} imply the following Monge-Amp\`{e}re equation for the entropy density,
\bea
\label{M-A entropy}
\cM(\sigma)=\partial_{n_x n_x} \sigma\ \partial_{n_y n_y} \sigma - \partial_{n_x n_y} \sigma\ \partial_{n_y n_x} \sigma= 1/\pi^2.
\eea
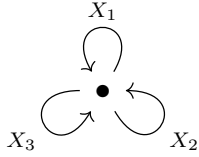
\begin{figure}
\centering
\begin{tikzcd}
\bullet \arrow[out=300,in=360,loop,swap, "X_2"]
  \arrow[out=60,in=120,loop,swap, "X_1"]
  \arrow[out=180,in=240,loop,swap, "X_3"]
\end{tikzcd}
\caption{Clover quiver diagram}
\label{clover}
\end{figure}
Having explained our general results for arbitrary toric quiver gauge theory, in the next section we explicitly compute the observables in some concrete examples.

\section{Some Examples}
\label{sec:examples}
In this section, we consider basic concrete examples of quivers and obtain analytic formulas for the profile function, the entropy density, phase structure, and BPS growth rate.
\subsection*{Clover Quiver\ $\mathbb{C}^3$}
\label{sec:clov}
The clover quiver, Fig.~\ref{clover} is characterized by Newton polynomial of $\mathbb{C}^3$ Calabi-Yau, $P(z,w)=1+z+w$. The Amoeba of this quiver is shown in Fig.~\ref{fig:Amoebas}.
The Following result for the Mahler measure of $P(z,w)=c+ az+ bw$, is essential in understanding the thermodynamics of the clover quiver.

\begin{figure}
\centering
\begin{tikzpicture}[scale=1.25]
\coordinate [label=left:$B$] (B) at (-1cm,-1.cm);
\coordinate [label=right:$A$] (A) at (2.2cm,-1.0cm);
\coordinate [label=above:$C$] (C) at (1cm,1.0cm);
\draw (A) -- node[below] {|c|} (B) -- node[above left] {|a|} (C) -- node[above right] {|b|} (A);
\pic[draw, -,"$\theta_1$",angle eccentricity=1.5] {angle=C--A--B};
\pic[draw, -,"$\theta_2$",angle eccentricity=1.5] {angle=A--B--C};
\pic[draw, -,"$\theta_3$",angle eccentricity=1.5] {angle=B--C--A};
\end{tikzpicture}
\caption{Triangle for clover quiver}
\label{triangle}
\end{figure}
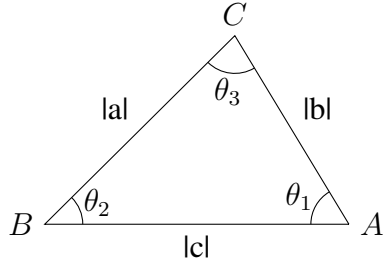

\begin{figure}
\centering
\label{fig:ent1}
\includegraphics[width=10cm]{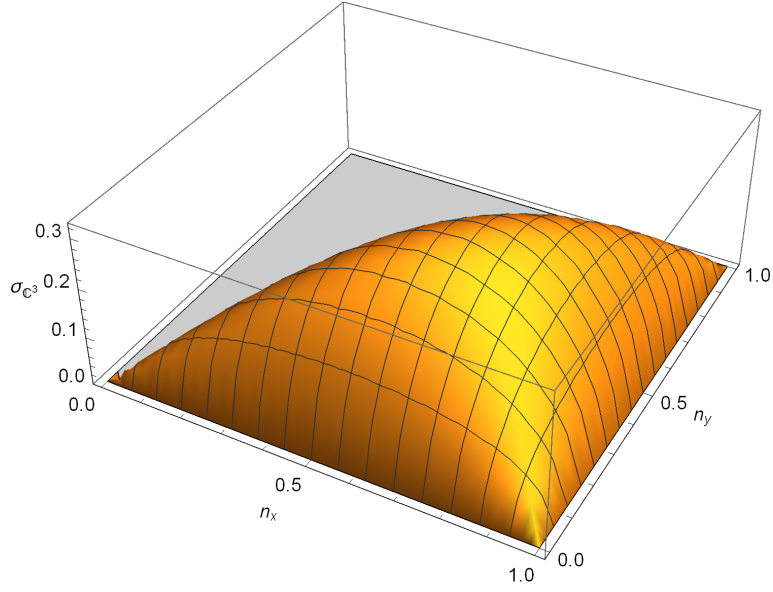}
\caption{Minus entropy density of clover quiver}
\end{figure}

\textbf{Theorem.} (Maillot) \cite{mail}. For $a,b,c \in \mathbb{C}^3$, and $(\log|a|, \log|b|, \log|c|)\in \cA_P$
\bea
\label{mah tri}
\pi m(c + az + bw)= D\left(\frac{|c|}{|a|}e^{\mathrm{i} \theta_2}\right)+\theta_1 \log |a|+ \theta_2 \log|b|+ \theta_3 \log |c|,
\eea
where $D$ is the Bloch-Wigner function, reviewed in Section~\ref{sec:LBW}. Outside the Amoeba, we have
\bea
m = \max (\log|a|,\log|b|,\log|c|).
\eea
$|a|,|b|,|c|$ can be seen as sides of a triangle, Fig.~\ref{triangle}, and Mahler measure in this case is given by Eq.~\eqref{mah tri}.
The profile of the clover quiver is the Ronkin function of  $\mathbb{C}^3$ and by using Eq.~\eqref{Ronkin}, it can be obtained from the Mahler measure in "Maillot Theorem" by putting $a=e^x$, $b=e^y$ and $c=1$. For $x,y \in \cA_{\mathbb{C}^3}$ ($|a|,|b|,|c| \in \Delta$), we obtain
\bea
\label{Ron 1}
N_{\mathbb{C}^3}(x,y)=m(1 + e^x z + e^y w)= \frac{1}{\pi}D(e^{-x+\mathrm{i} \theta_2})+\frac{\theta_1}{\pi} x+ \frac{\theta_2}{\pi} y,
\eea
and in unbounded complement of the Amoeba we have, 
\bea
N_{\mathbb{C}^3}(x,y)= \max(x, y, 0).
\eea
Comparing with Legendre transformation in Eq. \eqref{Leg}, we observe that
\bea
\label{ent den C}
\sigma_{\mathbb{C}^3}= -\frac{1}{\pi}D(e^{-x+\mathrm{i} \theta_2}), \quad n_x= \frac{\partial N_{\mathbb{C}^3}}{\partial x}=\frac{\theta_1}{\pi},\quad n_y=\frac{\partial N_{\mathbb{C}^3}}{\partial y}=  \frac{\theta_2}{\pi}.
\eea
Using the trigonometry of the triangle in Fig. \ref{triangle}, the slopes of $\mathbb{C}^3$ quiver is explicitly obtained as
\begin{equation}\label{eq: xderiv ronkin for hc}
n_x=\begin {cases}
0& \ x<\log|e^y-1|\\
\frac{1}{\pi}\cos^{-1}\left(\frac{e^{2y}-e^{2x}+1}{2e^y}\right)& \ \log|e^y-1| \leq x\leq\log|e^y+1|\\
1& \ x>\log|e^y+1|
\end {cases},
\end{equation}
and
\begin{equation}\label{eq: yderiv ronkin for hc}
n_y=\begin {cases}
0& \ y<\log|e^x-1|\\
\frac{1}{\pi}\cos^{-1}\left(\frac{e^{2x}-e^{2y}+1}{2e^x}\right)& \ \log|e^x-1| \leq y\leq\log|e^x+1|\\
1& \ y>\log|e^x+1|
\end {cases}.
\end{equation}
The slope can also be obtained directly by Residue computation of the derivatives of the Ronkin function \cite{Ya-Oo}.
\begin{figure}
\centering
\label{fig:Ron1}
\includegraphics[width=10cm]{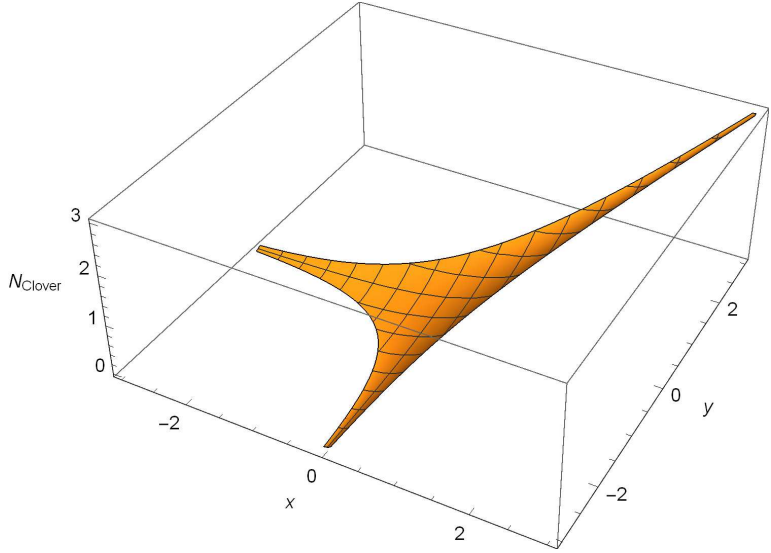}
\caption{Profile of clover quiver}
\end{figure}
Using Eqs. \eqref{Ron 1}, \eqref{eq: xderiv ronkin for hc} and \eqref{eq: yderiv ronkin for hc}, profile of the quiver, can be explicitly obtained in terms of $(x,y)$ coordinates,
\begin{eqnarray}
N_{\mathbb{C}^3}(x,y)&=& x\ n_x(x,y)+ y\ n_y(x,y) + \frac{1}{\pi} D(e^{-x+\mathrm{i}\pi n_y(x,y)}),
\end{eqnarray}
as it is plotted in Fig.~\ref{fig:Ron1}.
To compute the entropy density $\sigma_{\mathbb{C}^3}(n_x, n_y)$ as a function of slope, we need to solve the system of equations \eqref{eq: xderiv ronkin for hc} and \eqref{eq: yderiv ronkin for hc} for $x$, so we obtain
\bea
\label{inverse clover}
e^{-x}=\csc(\pi n_x)\sin\Big(\pi(n_x-n_y)\Big),
\eea
and then replacing the result in \eqref{ent den C}, we have
\begin{equation}
\sigma_{\mathbb{C}^3}(n_x,n_y)=-\frac{1}{\pi} D\bigg(e^{\mathrm{i} \pi n_y}\csc(\pi n_x)\sin\Big(\pi(n_x-n_y)\Big)\bigg).
\end{equation}
Slope of the dimer model is restricted to Newton polygon, $(n_x, n_y)\in NP(P)$, see Section \ref{sec:dimer}, and the entropy density in this region is plotted in Fig.~\ref{fig:ent1}. 

In order to explore the phase structure, we use a useful expression for the entropy density as function of slopes. 
Eq.~\eqref{ent clov loba} together with properties of the Bloch-Wigner function in Section~\ref{sec:LBW}, applied in Eq. \eqref{ent den C}, imply 
\bea
\label{ent lob1}
\sigma_{\mathbb{C}^3}(n_x,n_y)= -\frac{1}{\pi}\Big(L(\pi n_x) +  L(\pi n_y) +  L\left(\pi(1-n_x-n_y)\right)\Big).
\eea
This result, for the surface tension of the honeycomb dimer model is also obtained by other plausible methods in \cite{Ce-Ke}.

To compute BPS growth rate we need the Mahler measure $m_{\mathbb{C}^3}=m(1+z+w)$. This can be obtained from Eq.~\eqref{Ron 1}, by choosing $a=1$, $b=1$ and $c=1$ in Eq.~\eqref{mah tri}, or equivalently by the Ronkin function at origin
\bea
\label{clov mahler}
m_{\mathbb{C}^3}=N_{\mathbb{C}^3}(0,0)=-\sigma_{\mathbb{C}^3}(2/3,1/3)=\frac{1}{\pi}D(e^{\mathrm{i}\pi/3})=\frac{3}{\pi} L(\pi/3)\approx 0.32,
\eea
where the extremum point of the entropy density is obtained as $(n_x^*, n_y^*)=(2/3,1/3)$. This is consistent with the solution of $x=y=0$ in Eqs.~\eqref{eq: xderiv ronkin for hc} and \eqref{eq: yderiv ronkin for hc}, since for two opposite angles of triangle, $\theta_1$ and $\theta_2$, angle $\theta_1$ is $\pi-\theta_1$ in the unit circle.
Thus, the BPS growth rate in this quiver can be computed explicitly,
\bea
\label{clov growth}
\log{\Omega_{\mathbb{C}^3}(n)}\sim \ m_{\mathbb{C}^3}^{1/3} \ n^{2/3}\approx \ 0.68\ n^{2/3}.
\eea
This is in agreement with the asymptotics $ (\zeta(3)/4)^{1/3}\ n^{2/3}$ of the crystal model of $\mathbb{C}^3$ quiver, the plane partition.

\subsection*{Resolved Conifold Quiver \ $\mathcal{C}$}
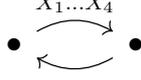
\begin{figure}
\centering
\begin{tikzcd}
\bullet  \arrow[r,bend left, "X_{1}... X_{4}"] & \bullet \arrow[l,bend left] 
\end{tikzcd}
\caption{Conifold quiver diagram}
\label{kronecker}
\end{figure}

Resolved conifold quiver, Fig.~\ref{kronecker}, is characterized by Newton polynomial $P(z,w)= e^{-t} z w + z + w -1$. 
The Amoeba of the resolved conifold is plotted in Fig.~\ref{fig:Amoebas}.
Statistical mechanics of the resolved conifold quiver is encoded in following result for the Mahler measure of $\cC$-type Calabi-Yau. 

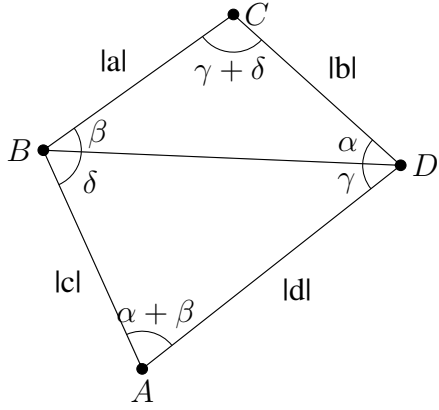
\begin{figure}
\begin{center}
\begin{tikzpicture}
\coordinate (B) at (-2.5,2.2);
\coordinate (A) at (-1.2,-.7);
\coordinate (D) at (2.2,2);
\coordinate (C) at (0,4);
 
\filldraw
(D) node[align=left, below] {}
-- (B) node[align=center, below, right] {}
(A) circle (2pt) node[align=left, below] {$A$}
-- node[below right] {|d|} (D) circle (2pt) node[align=center, below, right] {$D$}
-- node[above right] {|b|} (C) circle (2pt) node[align=left, above, right] {$C$}
-- node[above left] {|a|} (B) circle (2pt) node[align=left, above, left] {$B$}
-- node[below left] {|c|} (A);
\pic [draw, -, "$\alpha$", angle eccentricity=1.5] {angle = C--D--B};
\pic [draw, -, "$\beta$", angle eccentricity=1.5] {angle = D--B--C};
\pic [draw, -, "$\alpha+\beta$", angle eccentricity=1.5] {angle = D--A--B};
\pic [draw, -, "$\delta$", angle eccentricity=1.5] {angle = A--B--D};
\pic [draw, -, "$\gamma+\delta$", angle eccentricity=1.5] {angle = B--C--D};
\pic [draw, -, "$\gamma$", angle eccentricity=1.5] {angle = B--D--A};
\end{tikzpicture}
\caption{Cyclic quadrilateral for resolved conifold}
\label{fig:quad}
\end{center}
\end{figure}

\textbf{Theorem.} (Vandervelde) \cite{Van}. Consider Newton polynomial $P(z,w) = azw+ bz+cw+d$ for $a,b,c,d \in \mathbb{R}$. Suppose there is a non-degenerate convex cyclic quadrilateral $ABCD$ with edge lengths $|a|$, $|b|$, $|d|$, $|c|$ in cyclic order, inscribed in a circle and angles $\alpha$, $\beta$, $\delta$, $\gamma$, of arcs cut off by the edges of the quadrilateral, see Fig.~\ref{fig:quad}. Then, the Mahler measure, inside the Amoeba, is given by
\bea
\pi m= D(\frac{a u}{b}) - D(\frac{c u}{d}) + \alpha \log |a|+ \beta \log |b|+\gamma \log |c|+\delta \log |d|,
\eea
and in the non-quadrilateral case, outside Amoeba, we have
\bea
m = \max (\log|a|,\log|b|,\log|c|,\log|d|),
\eea
where $u$ is the solution for $w$ of the following equations,
\bea
\label{u}
P(z,w)=0,\quad |z|=1,\ |w|=1.
\eea

To obtain the profile of the resolved conifold quiver from Eq. \eqref{Ronkin}, we need to make the following choice $a=e^{-t+x+y}, b= e^x, c=e^y, d=-1$ in the above Mahler measure in "Vandervelde Theorem", and thus for $x,y$ inside the Ameoba we have
\bea
\label{Ronkin conif}
\pi N_{\cC_t}(x,y;t)= D(e^{-t+y} u_t)-D(-e^y u_t) + (\alpha+\beta) x + (\alpha+\gamma) y - \alpha t,
\eea
otherwise,
\bea
N_{\cC_t}(x,y;t) = \max (-t+x+y,x,y,0),
\eea
where we obtain $u$ from Eq. \eqref{u},
\bea
\label{resolved u}
u_t(x,y)=\frac{e^{-y}}{2(1+ e^{2x-t})}\Big(r+\sqrt{r^2-4(e^y+e^{y+2x-t})^2} \Big), 
\eea
with $r=1+e^{2y}-e^{2x}-e^{2x+2y-2t}$.
The angles $\alpha,\beta, \gamma, \delta$ of the cyclic quadrilateral can be find in terms of $x,y,t$, using our choice of the parameters $a,b,c,d$ and the trigonometry of the cyclic quadrilateral,
\bea
\label{quad trig}
\cos(\alpha+\beta)= -\frac{|a|^2+|b|^2-|c|^2-|d|^2}{2(|a||b|+|c||d|)},\quad \cos(\alpha+\gamma)= \frac{|b|^2+|d|^2-|a|^2-|c|^2}{2(|b||d|+|a||c|)},\nonumber
\eea
\bea
\cos(\beta+\delta)= \frac{|a|^2+|c|^2-|b|^2-|d|^2}{2(|a||c|+|b||d|)},\quad
\alpha+\beta+\gamma+\delta=\pi.
\eea
Comparing Eqs.~\eqref{Ronkin conif} and \eqref{Leg}, we obtain the entropy density and slope of resolved conifold quiver,
\bea
\label{conif ent}
\sigma_{\cC_t}=-\frac{1}{\pi}  \big( D(e^{-t+y} u_t)-D(-e^{y} u_t)  - \alpha t\big),
\eea
\bea
\label{slope con}
n_x=\frac{\partial N_{\cC_t}}{\partial x}= \frac{\alpha+\beta}{\pi}, \quad n_y=\frac{\partial N_{\cC_t}}{\partial y}=\frac{\alpha+\gamma}{\pi}.
\eea
By using Eqs.~\eqref{resolved u}, \eqref{quad trig} and \eqref{slope con}, one can make the following observation
\bea
u_t(x,y)=e^{\mathrm{i} \pi n_x}.
\eea
The entropy density can be written in terms of slope,
\bea
\label{conif ent1}
\sigma_{\cC_t}(n_x, n_y)= -\frac{1}{\pi} \Big(D\left(f(n_x,n_y)\right)-D\left(g(n_x,n_y)\right) - h(n_x, n_y) t\Big),
\eea
where $f(n_x,n_y)$, $g(n_x,n_y)$, and $h(n_x, n_y)$ are the inverse functions of  $e^{-t+y} u_t$, $-e^y u_t$, and $\alpha$, respectively. They can be obtained explicitly using Eq. \eqref{quad trig} with the choice of parameters for resolved conifolds and equations for slopes in Eq. \eqref{conif ent}.

Next we consider an important limit $t\rightarrow 0$ in the resolved conifold, to obtain the conifold quivers with $P(z,w)= -1+z+w+zw$. To obtain Profile function and entropy density of conifold, we take the limit  $t\rightarrow 0$ of Eq.~\eqref{Ronkin conif},
\bea
\label{con ent}
\pi N_{\cC}(x,y)&=&  D(e^{y} u_0) -D(-e^y u_0) + (\alpha'+\beta') x + (\alpha'+\gamma') y,\nonumber\\
\sigma_\cC&=& -\frac{1}{\pi} \Big( D(e^{y} u_0)-D(-e^{y} u_0) \Big).
\eea
where $\alpha', \beta', \gamma'$ are $t\rightarrow 0$ limit of $\alpha, \beta, \gamma$, and $u_0(x,y)=\lim_{t\rightarrow 0}u_t(x,y)$.

\begin{figure}
\centering
\includegraphics[width=10cm]{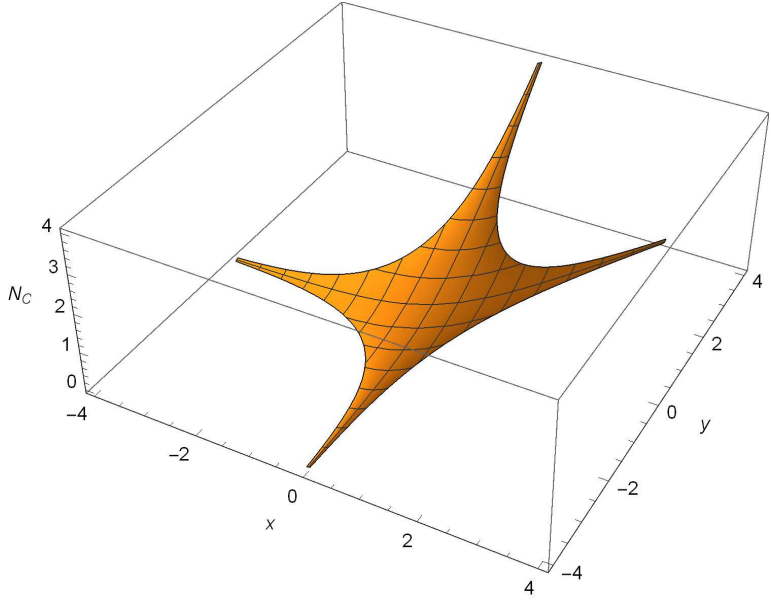}
\caption{Profile of conifold quiver}
\label{fig:conif prof}
\end{figure}

\begin{figure}
\centering
\includegraphics[width=10cm]{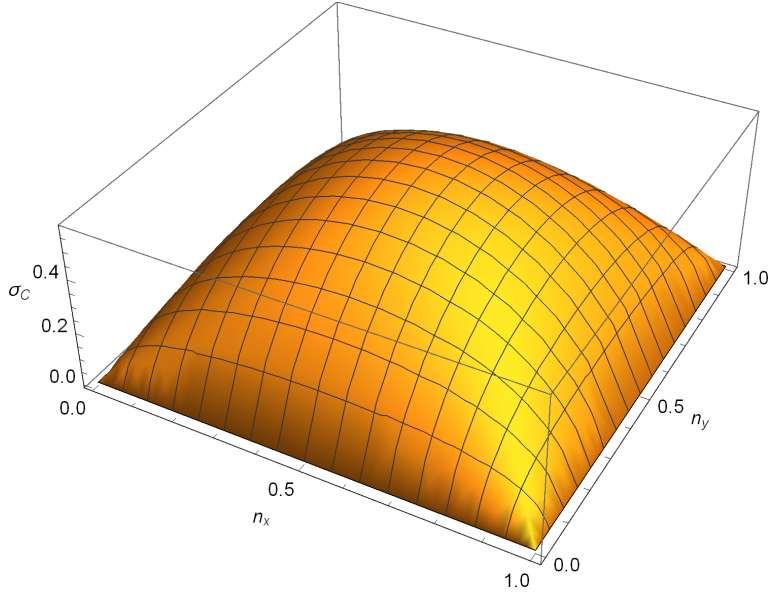}
\caption{Minus entropy density of conifold quiver}
\label{fig:conif ent}
\end{figure}

Solving Eqs.~\eqref{quad trig} for $y$, with the choice of conifold parameters $a=e^{x+y}, b=e^x, c=e^y, d=-1$, we obtain,
\bea
\label{con y}
e^y= \pm \sqrt{\csc^2{(\pi n_y)}\left(1-\cos{(2\pi n_x)}\cos^2{(\pi n_y)}+2 I\right)},
\eea
where $I=\sqrt{\cos^2{(\pi n_y)} \Big(1- \cos^2{(\pi n_x)}\cos^2{(\pi n_y)}\Big)\sin^2{(\pi n_x)}}$.
Using Eqs.~\eqref{con ent} and \eqref{con y}, we can write the entropy density in terms of the slopes of conifold,
\bea
\label{conif ent2}
\sigma_\cC= -\frac{1}{\pi} \Big(  D(e^{y} e^{\mathrm{i}\pi n_x})-D(e^{y} e^{\mathrm{i}\pi (n_x+1)}) \Big),
\eea
where $\pi n_x= \alpha'+ \beta'$ and $\pi n_y = \alpha'+\gamma'$.
Entropy density is plotted in Fig.~\ref{fig:conif ent}.
Furthermore, profile function can be written in terms of $x,y$ coordinates by using Eqs.~\eqref{resolved u} and  \eqref{quad trig} in the first line of Eq.~\eqref{con ent}. Profile function of the conifold quiver is plotted in Fig.~\ref{fig:conif prof}.

We can find the entropy density of conifold in terms of the Lobachevsky function. First, define $\xi=e^y u_0$, then by using properties of the Bloch-Wigner function and Lobachevsky function, from Eq. \eqref{con ent} we obtain 
\bea
\sigma_\cC&=& -\frac{1}{\pi} \Big(D(\xi) + D(1+\xi)\Big)\nonumber\\
&=& -\frac{1}{\pi} \Big(L(\arg(\xi))+ L(\arg(1-\frac{1}{\xi}))+ L(\arg(\frac{1}{1-\xi}))\nonumber\\
&&+L(\arg(-\frac{1}{\xi}))+ L(\arg(\frac{\xi}{1+\xi}))+L(\arg(1+\xi))\Big).
\eea
Remember that $L$ is an odd and $\pi$-periodic function, thus we have $L(\arg(\xi))+L(\arg(-\frac{1}{\xi}))=0$ and
\bea
\label{ent lobach}
\sigma_\cC &=& -\frac{1}{\pi} \left(L(\arg(1-\frac{1}{\xi}))+ L(\arg(\frac{1}{1-\xi}))+ L(\arg(\frac{\xi}{1+\xi}))+L(\arg(1+\xi))\right)\nonumber\\
&=&-\frac{1}{\pi} \Big( L(\alpha') +  L(\beta') + L(\gamma') + L(\delta')\Big)\nonumber\\
&=&-\frac{1}{\pi} \Big( L(\alpha') +  L(\beta') + L(\gamma') - L(\alpha'+\beta'+\gamma')\Big).
\eea
This result exactly matches with the following result in domino tiling \cite{Co-Ke},
\bea
\label{ent quadri}
\sigma_\cC= -\frac{1}{\pi} \sum_{i=1}^4 L(p_i),
\eea
where $p_i$ is the arc length on the circle cut off by $i$-th edge of the quadrilateral. 

Limit $t\rightarrow \infty$ is another interesting limit, in which we have $a\rightarrow 0$ and $\alpha\rightarrow 0$. Moreover, using the relation between angle and length of the chord on the unit circle, $\alpha=\arcsin{(\frac{e^{-t+x+y}}{2})}$, we obtain $\lim_{t\rightarrow\infty}\alpha t= \lim_{t\rightarrow\infty}\arcsin(\frac{e^{-t+x+y}}{2})t \rightarrow 0$. Thus, in this limit, using Eq.~\eqref{conif ent}, we obtain
\bea
\label{}
\lim_{t\rightarrow \infty}\sigma_{\cC_t}= \frac{1}{\pi}D(-\zeta),
\eea
where we defined $\zeta=\lim_{t\rightarrow{\infty}}e^{y} u_t(x,y)$, and from Eq.~\eqref{resolved u}
\bea
\lim_{t\rightarrow{\infty}}u_t(x,y)=\frac{1}{2}e^{-y}\Big(1+e^{2y}-e^{2x}+\sqrt{(-1-e^{2y}+e^{2x})^2-4e^{2y}} \Big).
\eea
In the resolved conifold, parameter $t$, for large values of $t$, can be seen as the length of the internal leg in the Amoeba of resolved conifold, Fig.~\ref{fig:Amoebas}. Amoeba of resolved conifold consists of two $\mathbb{C}^3$ vertices with a phase difference of $\pi$ and separated by an internal leg of length $t$. As $t\rightarrow\infty$ we have two isolated vertices. In this limit, the length of the internal leg of the Amoeba is infinitely large and, as we expect, we get the $\pi$-rotated  $\mathbb{C}^3$ vertex with entropy density $\sigma_{\pi(\mathbb{C}^3)}= \frac{1}{\pi}D(-\zeta)$ from Eq.~\eqref{ent clov loba}.
This is consistent with the entropy of clover quiver in Section \ref{sec:clov}, since we observe $\gamma=\theta_2, \delta=\theta_1$ in the limit $\alpha\rightarrow 0$.

\begin{figure}
    \centering
    \includegraphics[width=10cm]{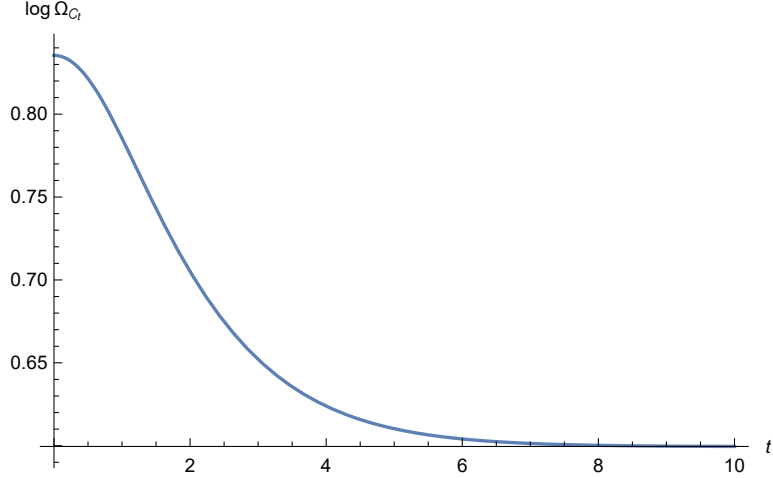}
    \caption{BPS growth rate of resolved conifold}
    \label{conifold rate}
\end{figure}

To study BPS growth rate of the resolved conifold, we need to find the Mahler measure with the following choice of parameters $a=e^{-t}, b=c=1, d=-1$. First, we find $u_{\cC_t}(t)$ by solving Eq.~\eqref{u},
\bea
u_{\cC_t}(t)= \frac{1 - e^{-2 t} + \sqrt{(-3 + e^{-t}) (1 + e^{-t})^3}}{2 + 2 e^{-t}}.
\eea
The dilogarithmic part of the Mahler measure of the resolved conifold can be obtained from Eq.~\eqref{Ronkin conif} by putting $x=0, y=0$ and removing linear term $-\alpha t$,
\bea
m_{\cC_t}(t)= \frac{1}{\pi} \Big( D(e^{-t} u_{\cC_t}(t))-D(-u_{\cC_t}(t)) \Big).
\eea
Then, associated BPS growth rate to this Mahler measure is given by
\bea
\log{\Omega_{\cC_t}(n)}\sim \ m_{\cC_t}^{1/3} \ n^{2/3},
\eea
and the numerical coefficient $m_{\cC_t}^{1/3}$ is plotted in Fig.~\ref{conifold rate}.

In the limit $t\rightarrow0$, we can obtain the Mahler measure of conifold, 
\bea
\lim_{t\rightarrow 0}(m_{\cC_t}(t)-\frac{\alpha}{\pi} t)=m_\cC= N_{\cC}(0,0)=-\sigma_{\cC}(1/2,1/2),
\eea
\bea
m_{\cC}= \frac{1}{\pi} \Big( D(\mathrm{i}) - D(-\mathrm{i})\Big)
=\frac{2}{\pi} D(e^{ \mathrm{i} \pi/2})= \frac{4}{\pi} L(\pi/2)=\frac{2}{\pi} \Im [Li_2(e^{ \mathrm{i} \pi/2})].
\eea
We observe the Catalan constant $G$ appears in this case,
\bea
m_{\cC}=\frac{2G}{\pi}\approx 0.58.
\eea
This result matches with the asymptotic number of domino tilings \cite{Kas, Fi-Te}.
Thus we obtain the BPS growth rate of conifold
\bea
\log{\Omega_{\cC}(n)}\sim \ m_{\cC}^{1/3} \ n^{2/3}= \ (\frac{2 G}{\pi})^{1/3}\ n^{2/3}\approx 0.83 \ n^{2/3}.
\eea

Mahler measure in the limit $t\rightarrow\infty$, becomes
\bea
\lim_{t\rightarrow \infty}m_{\cC_t}=-\frac{1}{\pi} D(e^{-2 \mathrm{i} \pi/3})= -\frac{1}{\pi} \Im [Li_2(e^{ -2\mathrm{i}\pi /3})]\approx 0.21.
\eea
Compare to Eq. \eqref{clov mahler}, we observe a rotation by $\pi$ in the clover quiver result,
\bea
\lim_{t\rightarrow\infty} m_{\cC_t}= -m_{\pi(\mathbb{C}^3)}=-\frac{1}{\pi} D(e^{\mathrm{i}(\pi+\pi/3)}).
\eea
This is consistent from the geometrical point of view, as we discussed before.
The BPS growth rate in the limit becomes
\bea
\lim_{t\rightarrow\infty}\log{\Omega_{\cC_t}(n)}\sim \ \lim_{t\rightarrow\infty} (m_{\cC_t})^{1/3}\ n^{2/3}
\approx 0.59\ n^{2/3}.
\eea
However, in limit $t\rightarrow \infty$, the term $e^{-t}zw$ in the Newton polynomial vanishes and we expect to reproduce the clover quiver result. This result can be recovered by a $\pi$-transformation ($-2\pi /3\rightarrow \pi- (-2\pi/3)=-\pi/3$) in our result. This is because, for a $\pi$-rotated $\mathbb{C}^3$ vertex, the appropriate angle to consider in the unit circle is obtained as $\pi-\theta$ instead of $\theta$.
Finally, twisted BPS growth rate in this case is
\bea
\lim_{t\rightarrow\infty}\log{\Omega^*_{\cC_t}(n)}\sim \ \big(-\frac{1}{\pi}D(e^{-\mathrm{i}\pi/3})\big)^{1/3}\ n^{2/3}
= m_{\mathbb{C}^3}^{1/3} \ n^{2/3}
\approx 0.68\ n^{2/3}.
\eea

\subsubsection*{Phase Structure and PDEs for Clover and Conifold Quivers}
\begin{figure}
\centering
\includegraphics[width=5cm]{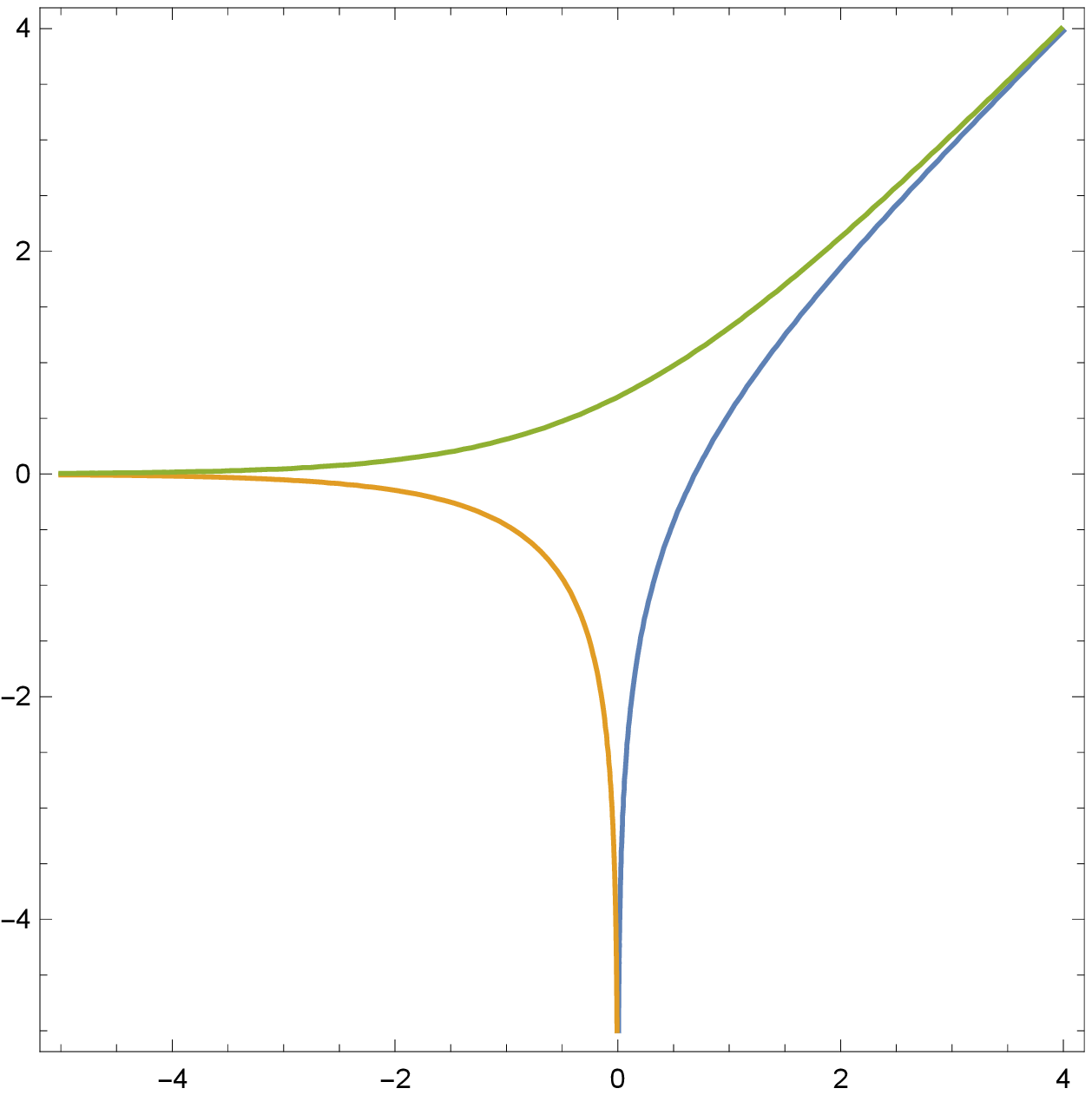}
\includegraphics[width=5cm]{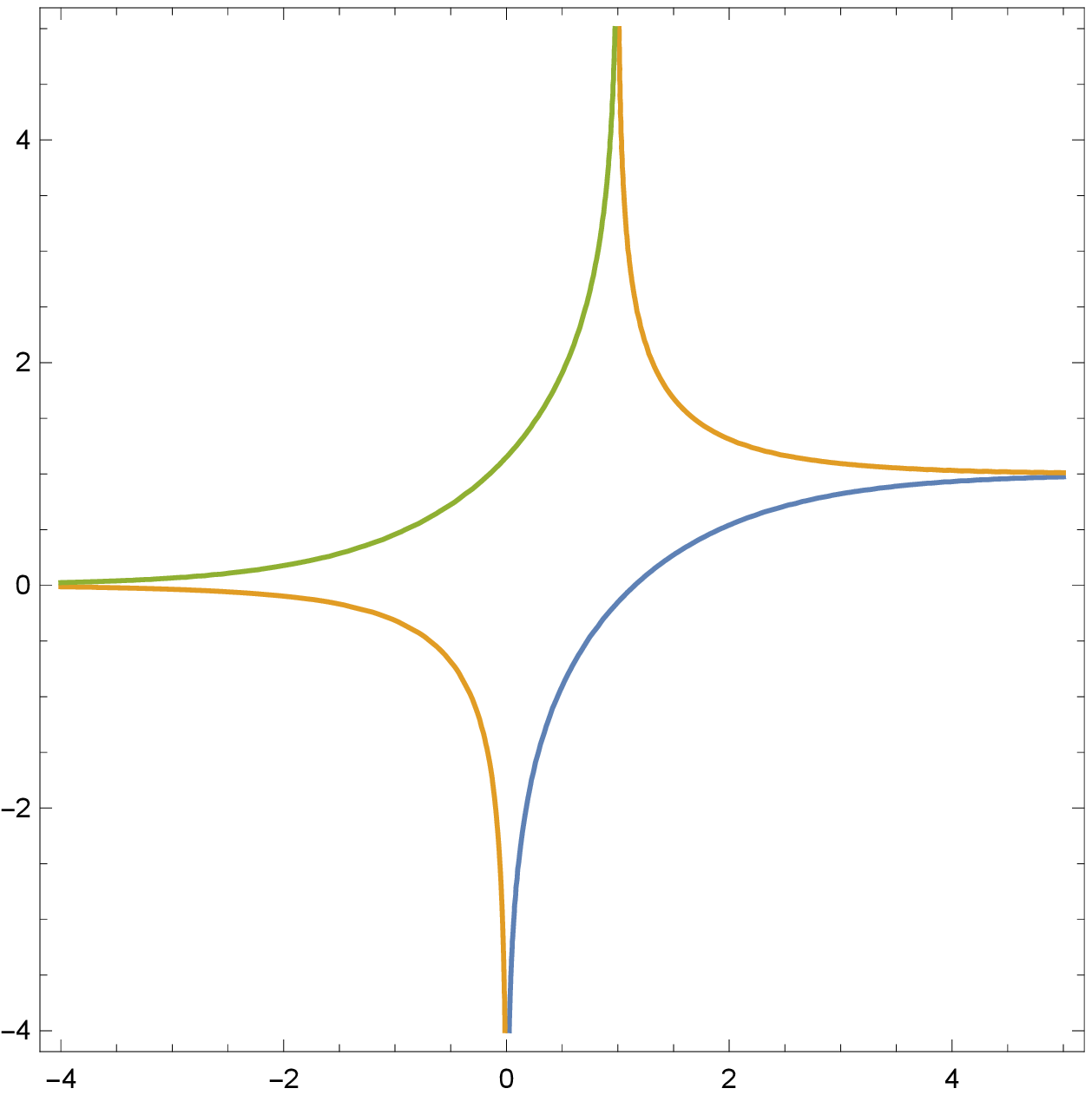}
\includegraphics[width=5cm]{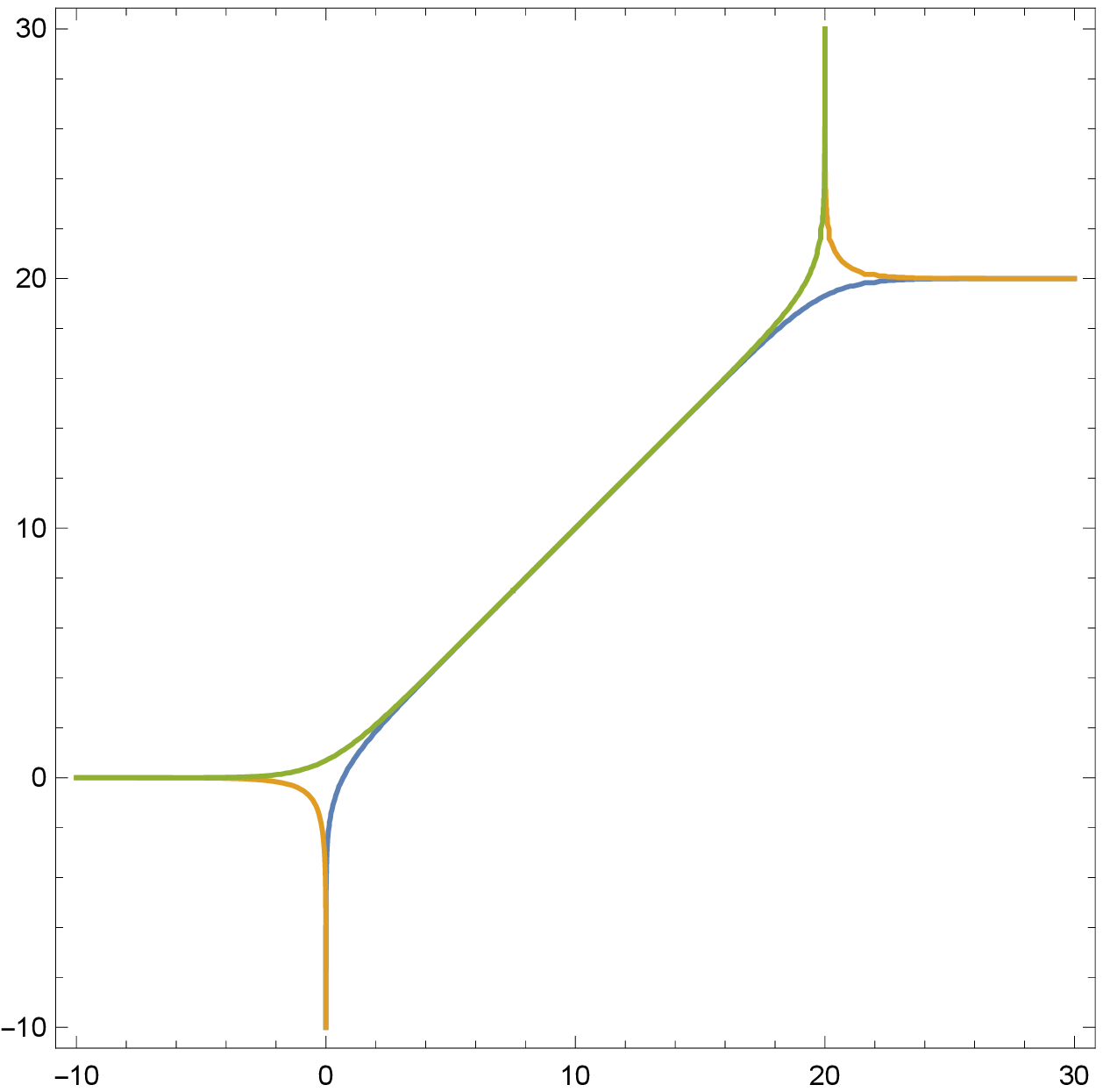}
\caption{Amoebas of clover (left), conifold (middle), and resolved conifold (right) quivers}
\label{fig:Amoebas}
\end{figure}

The phase structures of clover and conifold quivers are the Amoebas of associated Newton polynomials, plotted in \ref{fig:Amoebas}.
As we explained in Section~\ref{sec:phase}, zero slope limit, $n_x, n_y\rightarrow 0$ gives the boundary of the phase transition. We consider a facet of quiver, in which $(s,t)=(0,0)$ and study the continuity of entropy density and it derivative across the boundary of this facet. Using Eq. \eqref{ent lob1}, and the following approximation of Lobachevsky function near $\theta=0$,
\bea
L(\theta)\approx - \theta \log \theta + O(\theta^2), 
\eea
entropy density of clover quiver near the zero slope becomes
\bea
\sigma_{\mathbb{C}^3}(n_x, n_y)\approx n_x\log (\frac{n_x}{n_x+n_y})+n_y\log (\frac{n_y}{n_x+n_y})+ O(n^2_x, n^2_y).
\eea
Then, we observe that entropy density vanishes in the zero slope limit since $\lim_{n_x \rightarrow 0} \sigma(n_x,n_y)=\lim_{n_y \rightarrow 0} \sigma(n_x,n_y)=0$ and this confirm its continuity across the boundary of facet. However, we observe that the first derivative of entropy density in the limits $n_x\rightarrow 0$ and $n_y\rightarrow 0$ does not commute and in fact it diverges as $\lim_{n_x \rightarrow 0} \frac{\partial \sigma(n_x,n_y)}{\partial n_x}\rightarrow\infty$.
Thus, the first derivative of entropy density is not continuous on the boundary of the facet and we observe a phase transition. According to the discussion in Section \ref{sec:phase}, this result suggests a possible third order phase transition.

In the case of conifold quiver, the slope is given by Eq.~\eqref{slope con}, and the zero slope limit implies $\alpha,\beta,\gamma\rightarrow 0$ and $\delta\rightarrow \pi$. Thus, using Eq.~\eqref{ent quadri}, close to zero slope, in the leading order, we obtain
\bea
\sigma_{\cC}\approx \alpha\log \alpha +\beta\log\beta + \gamma\log \gamma.
\eea
Similar to clover quiver, the entropy density is continuous across the boundary of the zero slope facet but since the first derivative of  $\theta\log\theta$ term diverges at zero, we observe a discontinuity in the entropy density of conifold quiver, and thus a possible third-order phase transition. Considering the entropy density expressed as Bloch-Wigner function, the result of this section is consistent with the continuity but non-differentiability of the Bloch-Wigner function at the two points 0 and 1, mentioned in \cite{Zag}. A proper study of phase structure of the quivers considering all the facets of quiver requires further works and remains for future.

From the Monge-Amp\`{e}re equation \eqref{M-A entropy} in $\mathbb{C}^3$ quiver with entropy density in Eqs.~\eqref{ent den C} and \eqref{ent lob1}, we find an explicit solution for Monge-Amp\`{e}re equation,
\bea
\cM \Big( D(e^{-x+ \mathrm{i} \pi n_y})\Big)=\cM\Big(L(\theta_1) +  L(\theta_2) - L(\theta_1+\theta_2)\Big)=1/\pi.
\eea
Furthermore, by using Eqs.~\eqref{conif ent2} and \eqref{ent lobach} in the conifold quiver we have
\bea
\cM \Big( D(e^{y+ \mathrm{i} \pi n_x})-D(-e^{y+ \mathrm{i} \pi n_x})\Big)&=&\cM\Big(L(\alpha') +  L(\beta') + L(\gamma') - L(\alpha'+\beta'+\gamma'))\Big)\nonumber\\
&=&1/\pi
\eea
We make an interesting observation that the Lobachevsky and Bloch-Wigner functions are the solutions of Monge-Amp\`{e}re equation. We are not aware of any possible application or implications of this result for PDEs.
\begin{figure}
    \centering
    \includegraphics[width=5cm]{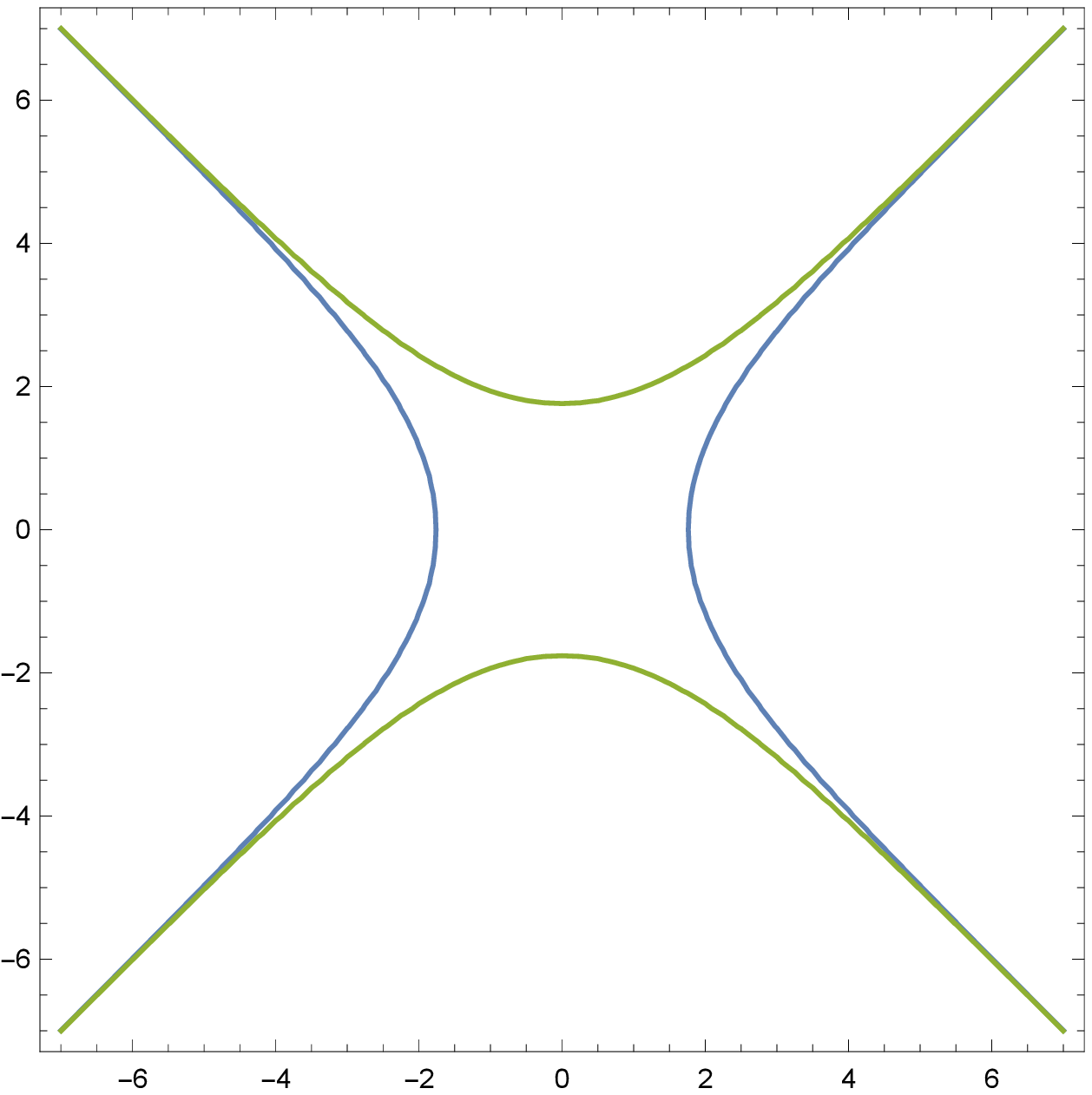}
    \includegraphics[width=5cm]{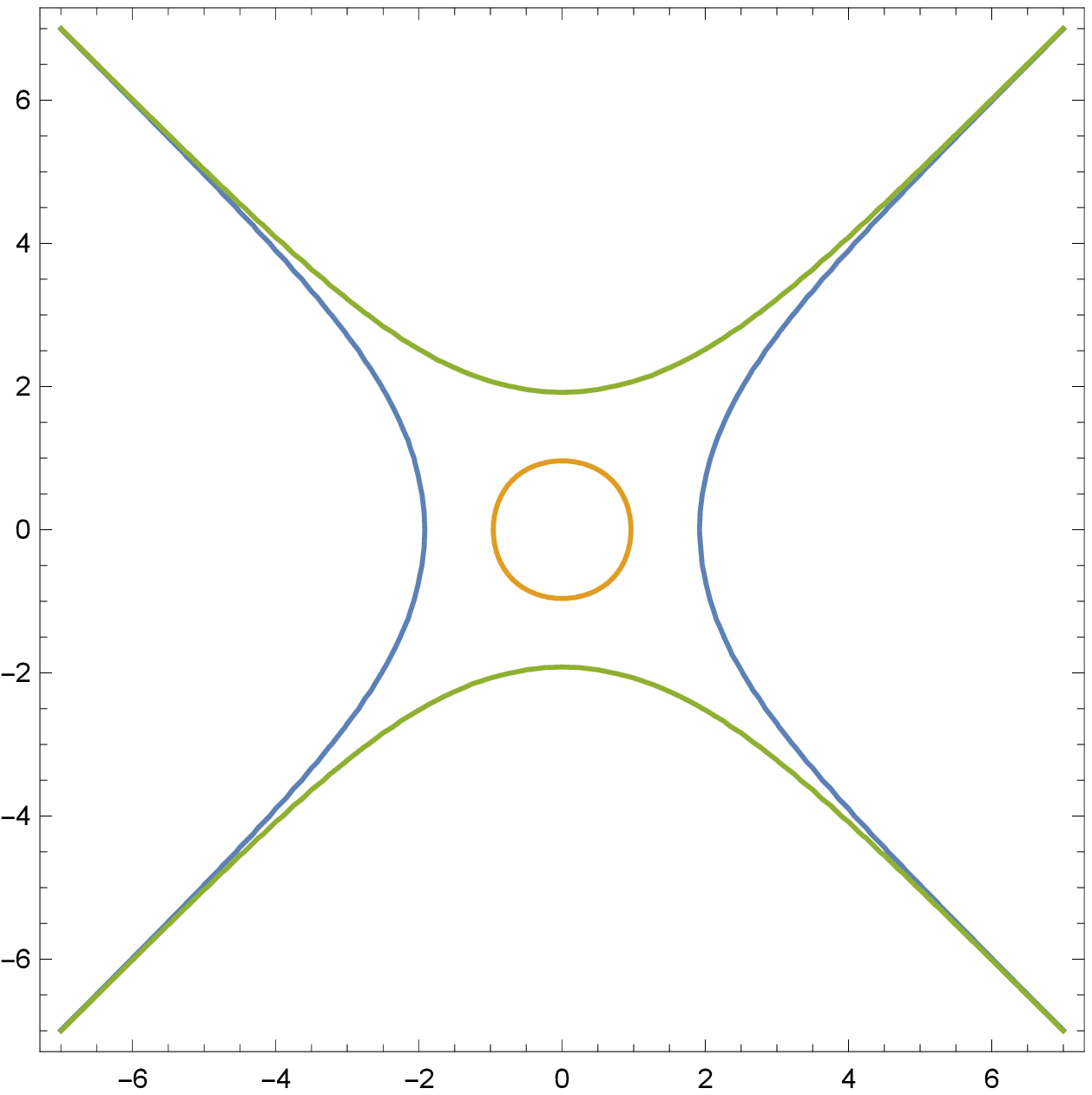}
    \caption{Amoebas of $ \mathbb{F}_0$ quiver at $k=4$ (left) and $k=5$ (right)}
    \label{AmoebaP}
\end{figure}

\subsection*{Hirzebruch Quiver\ $\mathbb{F}_0$}
Next example is $\mathbb{F}_0$ quiver characterized with Newton polynomial $P(z,w)=z+w+\frac{1}{z}+\frac{1}{w}-k$. This quiver is dual to the square-octagon dimer model \cite{Kenlec}. The Amoeba of this quiver is plotted in Fig.~\ref{AmoebaP} for two different values of parameter $k$. The bounded component of compliment of the Amoeba is the gas phase of the quiver and its area is a function of $k-4$. For $k>4$, this is the simplest quiver with a gaseous phase.

\begin{figure}
    \centering
    \includegraphics[width=10cm]{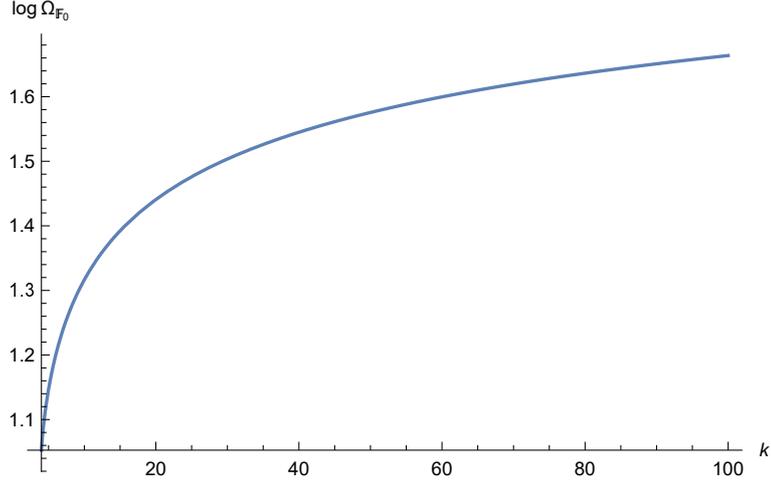}
    \caption{BPS growth rate of $\mathbb{F}_0$ quiver}
    \label{local P1 rate}
\end{figure}

The current techniques in Mahler measure theory do not allow to explicitly compute Mahler measure of $P(e^x z, e^y w)$. Thus, we can not develop statistical mechanics of this quiver to obtain the Ronkin function and entropy density of this quiver. However, we can obtain the BPS growth rate from the Mahler measure of $P(z,w)$ for $k>4$, computed in \cite{Vil},
\bea
\label{mah P1}
m_{\mathbb{F}_0}(k)=\frac{1}{2} \Bigg(\log k^{-2}-4k^{-2}\ {}_4 F_3\left(\begin{matrix}&3/2& &3/2& &1& &1&\\ \hspace{.1cm} &2&
&2& &2&\end{matrix};\ 16 k^{-2}\right)\Bigg),
\eea 
where $F$ is a hypergeometric function,
\bea
{}_p F_q\left(\begin{matrix}&a_1& &a_2& &...& &a_p&\\&b_1&
&b_2& &...&  &b_q&\end{matrix};\ x\right)=\sum_{n=0}^{\infty} \frac{(a_1)_n (a_2)_n ... (a_p)_n}{(b_1)_n (b_2)_n ... (b_q)_n}\frac{x^n}{n!},
\eea
and $ (c)_n= \frac{\Gamma(c+n)}{\Gamma(c)}$.
Using the Mahler measure, we can obtain the BPS growth rate in this quiver,
\bea
\log{\Omega_{\mathbb{F}_0}(k)}\sim \ m_{\mathbb{F}_0}(k)^{1/3} \ n^{2/3}.
\eea 
BPS growth rate dependence on $k$, for $k>4$, is plotted in Fig.~\ref{local P1 rate}. From Fig. \ref{local P1 rate}, we observe that BPS growth rate is increasing with the size of the gaseous phase, $k$. 
At $k=4$, one can show that
\bea
m_{\mathbb{F}_0}(4)=\frac{4 G}{\pi}= 4\pi^{-1} D(e^{\mathrm{i}\pi/2})= 4\pi^{-1} \Im [Li_2( \mathrm{i} \pi/2)]\approx 1.16,
\eea
and we observe $m_{\mathbb{F}_0}(4)=2 m_\cC$.
Finally, minimum of BPS growth rate is
\bea
\log{\Omega_{\mathbb{F}_0}(4)}\sim \ m_{\mathbb{F}_0}(4)^{1/3} \ n^{2/3}= \ (\frac{4G}{\pi} )^{1/3}\ n^{2/3}\approx 1.05 \ n^{2/3}.
\eea

\subsection*{ $\mathbb{C}^3/\mathbb{Z}_p \times \mathbb{Z}_p$ Orbifold Quivers and BPS Black Holes}
As a natural generalization of clover and conifold examples, we consider orbifold quivers \cite{Ha-Ke}.
The Mahler measure of orbifold $\mathbb{C}^3/\mathbb{Z}_p \times \mathbb{Z}_p$ for $p\geq 2$, denoted by $\cO_p$, can be written as
\bea
m_{\cO_p}=\int_{|z|=1}\int_{|w|=1} \log|1+\sum_{i=1}^p (z^i+w^i)+ \sum_{i=1}^{p-1} z^i w^{p-i}|\ \frac{dz}{z}\frac{dw}{w}. 
\eea

Current techniques in Mahler measure theory, do not allow for explicit computation of above integral for a general $p$, however, the simplest case $p=2$, $\mathbb{C}^3/ \mathbb{Z}_2\times \mathbb{Z}_2$ orbifold quiver with the Newton polynomial $P(z,w)= 1+z+z^2+w+w^2+zw$, is analytically computed \cite{Van2},
\bea
m_{\cO_2}=\frac{1}{\pi} \left(2D(\mathrm{i})- \frac{3}{4}D(e^{\frac{2\pi \mathrm{i}}{3}})\right)=\frac{1}{\pi}\left(2G-\frac{3}{4} \Im \big[Li_2( e^{\frac{2\pi \mathrm{i}}{3}})\big]\right)\approx 0.42.
\eea
Thus, we can obtain the BPS growth rate for this orbifold,
\bea
\log{\Omega_{\cO_2}}\sim \ m_{\cO_2}^{1/3} \ n^{2/3}\approx 0.74  \ n^{2/3}.
\eea

In the following we propose a physical application of our results for certain type of BPS black holes associated to the orbifolds of $\mathbb{C}^3$, constructed in \cite{He-Va}. We will interpret the BPS growth rate as the entropy of such black holes.
A bijection between crystal melting configurations and BPS black holes given by $D0$-, $D2$- and $D4$-branes wrapped on collapsed cycles on the large $p$ orbifolds $\mathbb{C}^3/\mathbb{Z}_p \times \mathbb{Z}_p$ is established in \cite{He-Va}. The spectrum of BPS black hole configurations are generated by geometric transitions of brane configurations and parametrized by crystal melting configurations. In the limit of large charges, a stable configuration in the orbifold will produce a local model for a BPS black hole in four dimensions. At low energies, these charge configurations are explained by quiver quantum mechanics. The profile of the partially melted crystal is inscribed in the ranks of gauge groups in the quiver. These ranks of the gauge groups determine D-brane charges.

The untwisted RR D0-brane charge of a black hole configuration realized by branes wrapped on cycles in the orbifolds. The crystal melting model for counting $D0$ brane bound states in a $D6$ brane filling $\mathbb{C}^3$, gives the RR $D0$-brane charges of the orbifold BPS black holes. Up to a factor $1/p^2$ this charge is given by the total number of atoms in the corresponding crystal melting model and thus the charge is equal to sum over the ranks of gauge groups in quiver quantum mechanics, $Q(BH)=\frac{1}{|\Gamma|}\sum_i N_i$ with $|\Gamma|=p^2$ for $\mathbb{C}^3/\mathbb{Z}_p\times \mathbb{Z}_p$. The $D0$ charge characterizes the energy of the ensemble of the black holes, $E_{BH}=|\Gamma|Q(BH)$. Therefore, the partition function of the black hole is given by crystal melting model partition function
\begin{eqnarray}\label{eq: BH PF}
\mathcal{Z}_{BH}=\sum_{BH}q^{|\Gamma|Q(BH)}=\sum_{Quiver}q^{\sum_iN_i}=\prod\limits_{k\geq 1}\frac{1}{(1-q^k)^k}.
\end{eqnarray}

The charge of the BPS black hole in the large $p$ limit can be obtained from the profile function of the quiver in the thermodynamic limit. Thus, Ronkin function can be interpreted as the charge density function of the heavy state of the black hole and the integral of charge density gives the charge of the black hole $Q^*=\frac{1}{p^2}\int N(x,y)\ dx\ dy$. In the thermodynamic limit, consider the statistical average of ranks of the gauge groups in the quiver. Then, the average charge of the black hole can be calculated as
\bea \label{eq: average charge}
<Q>&=&\sum_i\frac{1}{p^2}<N_i>=\frac{1}{p^2}q\frac{\partial}{\partial q}\log \mathcal{Z}_{BH}=\frac{1}{p^2}\sum_{k=1}^{\infty}\frac{k^2q^k}{1-q^k}\nonumber\\
&\sim& \frac{1}{p^2 g_s^3}\int_{0}^{\infty}du\ \frac{u^2 e^{-u}}{1-e^{-u}}=\frac{2\zeta(3)}{p^2 g_s^3}.
\eea
Since $var(Q)\sim O(g_s^{-4})\rightarrow 0$, as $g_s\rightarrow 0$ we have $<Q>=Q^*$. Therefore, the black hole states in the thermodynamic limit tend to the typical state corresponding to limit shape and this typical state is in fact the state with the average charge.
Similarly, the entropy density of clover quiver can be interpreted as the entropy density of the black hole with a certain charge and the Mahler measure as the BH growth rate, up to a scaling power. Interpreting the BH growth rate as the entropy of the black hole, using Eqs.~\eqref{BPS growth2}, \eqref{clov growth} and \eqref{eq: average charge} we obtain
\bea
S_{BH}=\log \Omega_{BH}=\lim_{p\rightarrow\infty}(p^2 Q^*)^{\frac{2}{3}} m_{\mathbb{C}^3}^{\frac{1}{3}} = \frac{\zeta(3)}{ g_s^2}.
\eea

\section{Conclusion and Discussion}
In this work, we used the toroidal dimer model and Mahler measure theory to obtain the asymptotic observables of the toric quiver gauge theories in terms of Mahler measure. Our focus in this work was on explicit computations of the observales in concrete examples. As a result of these examples we developed a new statistical mechanics of the resolved conifold quiver.

The initial goal of this work was to explicitly obtain the numerical values of BPS growth rates in different examples of quivers. We found that BPS growth rate is proportional to Mahler measure to a certain scaling power, however, the factor of proportionality is not yet explicitly obtained. we conjecture that this factor is $1/c$ where $c$ is the number of different types of dimer in a specific dimer model. For example, in honey-comb dimer model we have three different types of dimers and with the factor of one-third inserted in Eq.~\eqref{clov growth}, it exactly matches with the asymptotic number of plane partitions. Following this conjecture and using obtained results for Mahler measures, we can compare the asymptotic BPS index of different quivers,
\bea
\Omega_{\mathbb{F}_0}> \Omega_\cC > \Omega_{\cO_2}> \Omega_{\mathbb{C}^3}.
\eea

Explicit computations of the profile function, entropy density and BPS growth rate in other examples of the toric quivers such as orbifolds of $\mathbb{C}^3$ and $\cC$ are highly interesting. A related direction for future studies is the full asymptotic analysis of $\mathbb{F}_0$ quiver in which the Amoeba has a gas phase and in principle we can count D6-D4-D2-D0 bound states. Analytic computations of the profile function and entropy density are not known in this example.

The class of isoradial dimer model is of special interest because of their relations to $\cN=1$ superconformal field theories and brane tilings \cite{Ha-Ve}. The surface tension and Mahler measure of isoradial dimer model and their relation to hyperbolic volumes of certain polyhedrons have been partially studied recently \cite{Ken-iso}, \cite{Lal-hyper}. In fact, the entropy density of clover and conifold quivers, obtained in this paper, are the Hyperbolic volumes of some ideal tetrahedron and pyramid. Further understanding of the connection between entropy density in this work or topological indices of $\cN=1$ quiver theory with hyperbolic volumes using techniques of dimer model is a possible continuation of this work.    

In this work, we mainly considered the leading order asymptotics to the BPS counting problem. The fluctuations around the Ronkin function contribute to higher order corrections. These corrections are presumably capture the partition function of higher genus topological strings.
Another interesting direction for future studies is to develop the asymptotic analysis for the D4-D2-D0 BPS bound states and corresponding two-dimensional crystal model \cite{Nish}. A plausible approach would be to consider the zero slope limit of the results of the three-dimensional crystal model.

\section{Acknowledgement}
I am deeply grateful to R. Szabo for collaboration in the early stages of this project and valuable discussions. I am thankful to A. Cazzaniga for useful discussions. 
This research is supported by research funds from the Centre for Research in String Theory, School of Physics and Astronomy, Queen Mary University of London, and National Institute for Theoretical Physics, School of Physics and Mandelstam Institute for Theoretical Physics, University of the Witwatersrand.

\bibliographystyle{plain}
\bibliography{references}

\end{document}